\documentclass[aps,prl,nofootinbib,twocolumn,floatfix,
letterpaper,superscriptaddress,showpacs]{revtex4}
\usepackage{graphicx}\usepackage{natbib}\usepackage{amsmath}
\begin{document}
 
\title{TeV Gamma Rays from Geminga and the Origin of the GeV Positron Excess}

\author{Hasan Y{\"u}ksel}
\affiliation{Bartol$\,$Research$\,$Institute$\,$and$\,$Department$\,$of$\,$Physics$\,$and$\,$Astronomy, University$\,$of$\,$Delaware,$\,$Newark,$\,$Delaware$\,$19716}

\author{Matthew D. Kistler}
\affiliation{Center for Cosmology and Astro-Particle Physics and Department of Physics, Ohio State University, Columbus, Ohio 43210}

\author{Todor Stanev}
\affiliation{Bartol$\,$Research$\,$Institute$\,$and$\,$Department$\,$of$\,$Physics$\,$and$\,$Astronomy, University$\,$of$\,$Delaware,$\,$Newark,$\,$Delaware$\,$19716}

\date{May 5, 2009}

\begin{abstract}
The Geminga pulsar has long been one of the most intriguing MeV--GeV gamma-ray point sources.  We examine the implications of the recent Milagro detection of extended, multi-TeV gamma-ray emission from Geminga, finding that this reveals the existence of an ancient, powerful cosmic-ray accelerator that can plausibly account for the multi-GeV positron excess that has evaded explanation.  We explore a number of testable predictions for gamma-ray and electron/positron experiments (up to $\sim 100$~TeV) that can confirm the first ``direct'' detection of a cosmic-ray source.
\end{abstract}

\maketitle

\textit{Introduction.}---
Geminga holds a place of distinction among gamma-ray sources, being the first pulsar to be discovered through gamma rays, with a history of observations through a variety of techniques~\cite{Bignami:1996tp}.  While one of the brightest MeV--GeV gamma-ray point sources in the sky, there was no certain evidence of high-energy activity beyond the immediate neighborhood of the pulsar or its x-ray pulsar wind nebula (PWN) until the recent detection by Milagro of gamma rays at $\sim\,$20 TeV from a region of $\sim\,$3$^\circ$ around the pulsar~\cite{Abdo:2007ad,Abdo:2009ku}.  This detection places Geminga among the growing class of TeV PWNe (e.g., \cite{Aharonian:2006zb,Aharonian:2006xx}) and is important for understanding aged pulsars and their winds.  An immediate consequence is the existence of a population of high-energy particles.

\begin{figure}[b!]
\includegraphics[width=\columnwidth,clip=true]{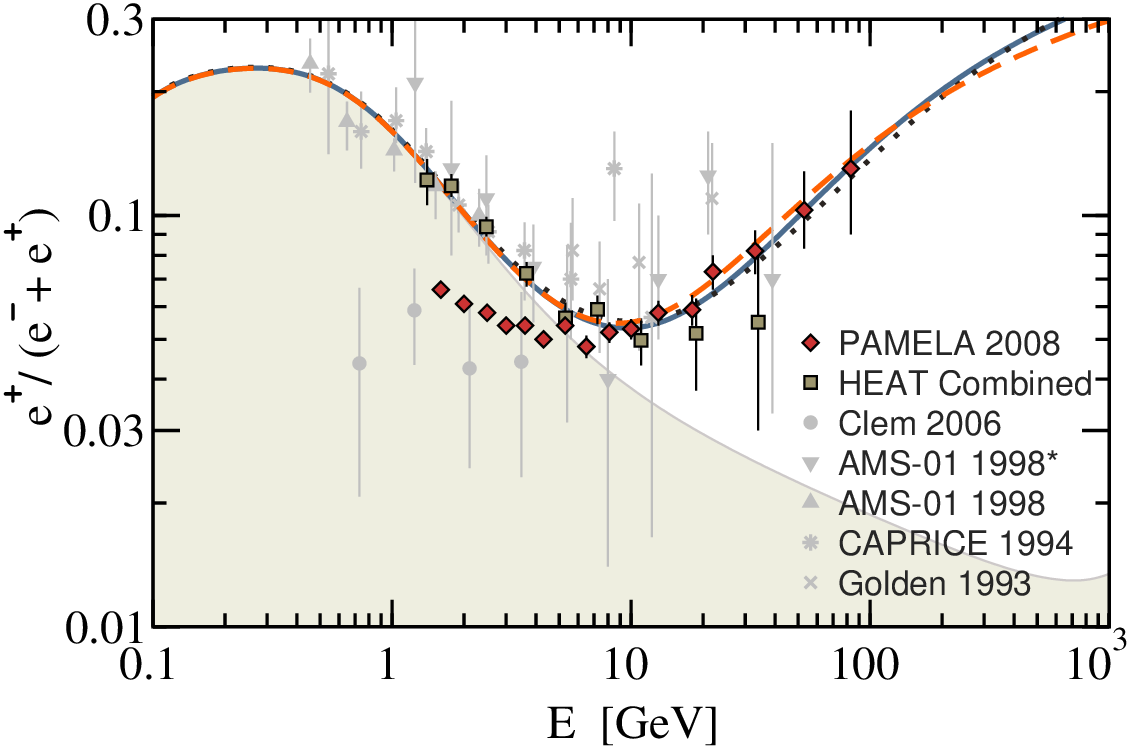}
\caption{The cosmic-ray positron fraction.  Shown are data compiled from Refs.~\cite{Beatty:2004cy,Adriani:2008zr,Aguilar:2007yf,clempriv:2008}, and scenarios based on the secondary model of Ref.~\cite{Moskalenko:1997gh} (shaded) and a plausible Geminga contribution (solid, dashed, and dotted lines) dependent upon distance and energetics (see text for details).
\label{excess}}
\end{figure}

The relative proximity of Geminga raises an interesting possibility, namely that these high-energy particles, most likely electrons and positrons, may be at the root of the explanation of the ``positron excess'', the observed~\cite{Barwick:1997ig,Beatty:2004cy,Adriani:2008zr} overabundance of multi-GeV positrons as compared to theoretical expectations~\cite{Moskalenko:1997gh} (see Fig.~\ref{excess}).  Severe energy losses of high-energy positrons require a local source of some kind~\cite{Coutu:1999ws}, such as Geminga~\cite{Aharonian(1995)} or even dark matter through its annihilation products~\cite{DM}.

Here, we connect the Milagro TeV gamma-ray ``halo'' to electrons and positrons with energies up to at least 100~TeV, expected to be accelerated in PWNe (e.g., \cite{Goldreich:1969sb,Rees:1974nr}; for a review see~\cite{Gaensler:2006ua}), and present several predictions.  Principally, while Geminga is apparently young enough to still produce high-energy particles, it is old enough that multi-GeV electrons and positrons from its more active past could have made it to Earth. The extended gamma-ray emission is strong evidence for $e^\pm$ production, acceleration, and escape, suggesting an explanation of the positron excess.  Moreover, this single nearest high-energy astrophysical source can reasonably account for the $e^- + e^+$ spectrum as measured by Fermi~\cite{Fermi:2009zk} and HESS~\cite{Aharonian:2008aaa,Aharonian:2009ah} with an extension to energies beyond several TeV, where no signal might be expected otherwise.

\textit{The Gamma-ray Source Next Door.}---
The observation of high-energy gamma rays from an astrophysical source implies the presence of higher-energy particles, typically $e^\pm$ or protons, that gave rise to them.  One striking element of the observation of $\sim\,$20 TeV gamma rays (with a significance of $4.9 \,\sigma$ in the PSF-smoothed map~\cite{Abdo:2007ad}, $6.3 \,\sigma$ for an extended source~\cite{Abdo:2009ku}) from Geminga by Milagro is the extent of the emission, $\theta\sim\,$3$^\circ$~\cite{Abdo:2007ad}, which corresponds to a physical size of $s_G \sim 10 \,$pc$ \,(\theta_G/3^\circ)(r_{G}/200\,{\rm pc})$, where $r_{G}$ is the distance to Geminga.  Since the angular resolution of Milagro is better than a degree and the characteristic age of the pulsar, $t_G \sim 3 \times 10^5\,$yr~\cite{Taylor:1993ba}, seemingly excludes a typical TeV supernova remnant, we shall consider an extended PWN with emission from a much larger region than seen in x-rays~\cite{Caraveo,Pavlov}.  We will draw guidance from the TeV-PWN HESS J1825--137~\cite{Aharonian:2006zb}, which, while only a tenth the age of Geminga, would appear tens of degrees wide if placed at $r_{G}\sim 200\,{\rm pc}$.

\begin{figure}[b!]
\includegraphics[width=\columnwidth,clip=true]{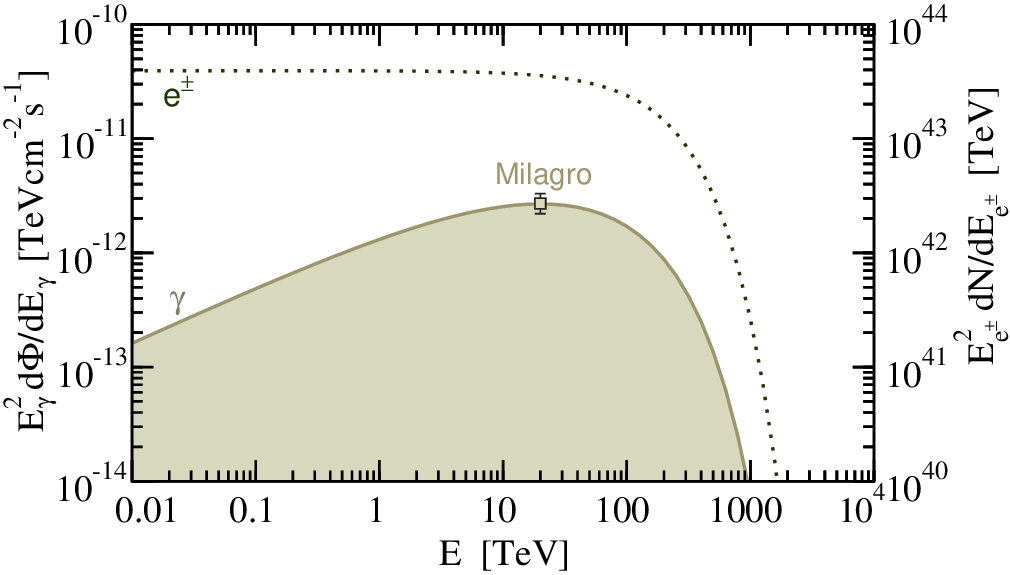}
\caption{Minimal inverse-Compton gamma-ray spectrum of the {\it extended} emission from Geminga (shaded) and the Milagro measurement at 20~TeV ({\it left axis}).  Also, the energy distribution (dotted line) of the associated $e^\pm$ ({\it right axis}).
\label{geminga}}
\end{figure}

We first examine whether the gamma rays can be explained through inverse-Compton (IC) up-scattering of cosmic microwave background (CMB) photons by $e^\pm$.  Note that the pulsar age exceeds the IC cooling time on CMB photons of the $\gtrsim\,$100~TeV $e^\pm$ needed to produce $\gtrsim\,$20~TeV gamma rays, $\tau_{IC}\sim 10^4 (100\,$TeV$/E_e)\,$yr in the Thomson limit.  Including synchrotron losses further decreases $\tau_{\rm cool}$, implying fresh $e^\pm$ production.  To account for the Milagro measurement of $6.9 \pm 1.6 \times 10^{-15}$ TeV$^{-1}$ cm$^{-2}$ s$^{-1}$ at $20$~TeV (see Fig.~\ref{geminga}), we consider a generic parent $e^\pm$ spectrum of the form $dN/d\gamma \propto \gamma^{-\alpha} e^{-\gamma/\gamma_{\rm max}}$, with $\gamma = E/ (m_e c^2)$.  Lacking more detailed observations, we choose $E_{\rm min}=1\,$~GeV, $E_{\rm max}=200\,$~TeV and $\alpha=2$ (typical to shock acceleration and as inferred in the Vela~X PWN~\cite{Aharonian:2006xx}).  The resulting IC spectrum is
\begin{equation}
\frac{d\Phi}{dE_\gamma} \! = \! \frac{c}{4 \pi r_G^2} \! \int \! d\gamma \! \int \!
   dE_{b} \frac{dN}{d\gamma} \, n_{b}(E_{b}) \, \sigma_{\rm KN}(\gamma,E_{b},E_\gamma),
\label{eq:icpoint}
\end{equation}
where the Klein-Nishina cross section, $\sigma_{\rm KN}$, is given by Ref.~\cite{Blumenthal:1970gc}.  The dominant scattering background, $n_{b}(E_{b})$, in the Milagro energy range is the CMB, allowing us to construct the minimal spectrum shown in Fig.~\ref{geminga} (at lower energies, other contributions become relevant).  When we normalize the IC spectrum to the Milagro TeV gamma-ray luminosity of $\dot {\cal E}_{\gamma{\rm , TeV}} \sim 10^{32}$~erg~s$^{-1}$ (for $r_{G}\sim 200\,{\rm pc}$), at least ${\cal E}_{e^\pm} \sim 10^{45}$~erg of $e^\pm$ is required.

We can relate this phenomenological spectrum to the pulsar.  The Goldreich-Julian flux~\cite{Goldreich:1969sb} is $\dot N_{\rm GJ} \simeq {B \, \Omega^2 \, R^3}/ e c$.  For Geminga, $R  \simeq 10$~km, $B \simeq 1.6 \times 10^{12}$~G, and $\Omega \simeq 26.5$~s$^{-1}$~\cite{Taylor:1993ba}, which yield $\dot N_{\rm GJ} \simeq 10^{32}$~s$^{-1}$.  For our $E_{e^\pm}^{-2}$ spectrum, $\left\langle E_{e^\pm} \right\rangle \approx 15$~GeV.
The total power is then $\dot {\cal E}_{e^\pm} \approx \dot N_{e^\pm} \left\langle E_{e^\pm} \right\rangle \approx \mathcal{M}\, \dot N_{\rm GJ} \left\langle E_{e^\pm} \right\rangle \approx 2 \times 10^{30}\,\mathcal{M}\,$erg~s$^{-1}$, where $\mathcal{M} = \dot N_{e^\pm} / \dot N_{\rm GJ}$ is the pair multiplicity.  For a pure electron flow ($\mathcal{M} \simeq 1$), the power beyond 10~TeV is only $\sim 5 \times  10^{29}$~erg~s$^{-1}$~$\ll \dot {\cal E}_{\gamma{\rm , TeV}}$, thus requiring pair production resulting in $\mathcal{M} \gtrsim 100$.  Including synchrotron losses comparable to IC increases the required $\mathcal{M}$.  This can only be avoided by assuming a spectrum much-harder than $E_{e^\pm}^{-2}$ or $E_{\rm min}>100\,$~GeV (both inconsistent with, e.g.,  Vela~X~\cite{Aharonian:2006xx,deJager2007}).

While we have been trying to explain the {\it observed} signal alone, there may well be more lower surface brightness emission at larger angles, and also at lower energies.  Two considerations make this likely.  First, the pulsar's $e^\pm$ output was probably much stronger in the past when its spin-down power was higher.  Secondly, no evidence of a large-scale radio or x-ray nebula exists, and a substantial fraction of the $e^\pm$ may be escaping, so that the above multiplicity is only a lower limit.  $\mathcal{M} \sim 10^4$ (as inferred for younger TeV PWNe~\cite{deJager2007} or pulsar models~\cite{Zhang:1999ua}) does not exceed the spin-down power of $\sim 10^{34.5}$~erg~s$^{-1}$ \cite{Taylor:1993ba}, so that a large pair conversion fraction is possible.

\textit{The Origin of the Positron Excess.}---
The confirmed presence of a nearby, ancient source of high-energy electrons and positrons immediately suggests an explanation for the positron excess.  If so, then we would essentially be living ``within'' the extended halo of the source, seeing $e^\pm$ that were accelerated long ago when the pulsar was stronger.  The density of particles at a given time and place, $n(r,t,\gamma)$, is governed by the diffusion equation, which (in spherically symmetric geometry) is
\begin{equation}
\frac{\partial \, n}{\partial t}= \frac{\mathcal{D}(\gamma)}{r^2} \frac{\partial}{\partial r}r^2  \frac{\partial \, n}{\partial r} + \frac{\partial}{\partial \gamma} [\ell(\gamma) \, n] +Q(\gamma)\,,
\label{eq:diff}
\end{equation}
where $Q$ is the source term.  In our range of interest, the energy loss rate is well approximated by $\ell(\gamma)= \ell_0 \gamma^2$. Considering IC losses on the CMB (energy density $\sim 0.3$~eV$\,$cm$^{-3}$) and synchrotron losses due to a $\sim 3\, \mu$G magnetic field ($\sim 0.2$~eV~cm$^{-3}$) yields $\ell_0\simeq 5 \times 10^{-20}$~s$^{-1}$.

Calculations that assess the positron flux from dark matter annihilation typically assume steady-state conditions (i.e., ignoring the time-dependent term).  However, our scenario is manifestly dynamic, with both the source luminosity and distance potentially changing with time.  The analytic solution given by Atoyan, Aharonian, and V\"{o}lk~\cite{Atoyan:1995} allows for this to be conveniently handled.  Their diffusion coefficient is $\mathcal{D}(\gamma)= \mathcal{D}_0 (1+\gamma/\gamma_*)^\delta$, with $\gamma_*\simeq 6 \times 10^3$~\cite{Atoyan:1995}.  In the limit of a single burst from a point source, $Q(\gamma) = dN/d\gamma \, \delta(r) \, \delta(t-t_G)$, with
\begin{equation}
n(r,t,\gamma)=\frac{dN/d\gamma}{\pi^{3/2}\,r^3} \frac{[{r}/{r_{d}(t,\gamma)}]^3 \, e^{-[{r}/{r_{d}(t,\gamma)}]^2}}{(1-\ell_0\, t\, \gamma)^{2-\alpha}}\, .
\label{eq:difsolburst}
\end{equation}
where the spectrum is cut off at $\gamma_{c}=1/(\ell_0 \, t)$.  The energy loss rate and age then set the maximal energy of particles that reach us today, with a diffusion radius of
\begin{equation} 
r_{d}(t,\gamma)\simeq 2\left(\mathcal{D}(\gamma) \, t \,
[{1-(1-\gamma/ \gamma_{c})^{1-\delta} }]/[{(1-\delta)\gamma / \gamma_{c}}]  \right)^{1/2}. \nonumber
\label{eq:difrad}
\end{equation}
For $t_G \sim 3 \times 10^5\,$yr, $\mathcal{D}_0\simeq 4 \times 10^{27}$~cm$^2$s$^{-1}$, and $\delta = 0.4$ (intermediate between $\delta$ = 1/3 and 1/2~\cite{Strong:2007nh}), the diffusion radius is $r_{d} \simeq 150,\, 175,\, 250$~pc for $E=2,10,50$~GeV particles, respectively, so multi-GeV particles are now arriving.  We caution against extrapolating to small radius, since the diffusion solution may transition~\cite{Aloisio:2008tx} to a wind-like $n\propto r^{-2}$ form (as does HESS J1825--137~\cite{Aharonian:2006zb}) near the source, although the data allow no firm conclusions.

For a continuously emitting source such as Geminga, the injection rate can be parametrized as
$d\dot{N}/d\gamma \propto {\cal L}_{e^\pm}(t) \gamma^{-\alpha} e^{-\gamma/\gamma_{\rm max}}$, with
${\cal L}_{e^\pm}$ the $e^\pm$ luminosity.  The local particle density is $n_{\odot}(\gamma)= \int^{t_{G}} dt\,  n(r_{G},t,\gamma)$.  Assuming braking via magnetic dipole radiation, the spin-down luminosity evolves as $\propto (1+t/t_0)^{-2}$~\cite{Gunn:1969ej}, with a pulsar-dependent timescale, $t_0$, and ${\cal L}_{e^\pm}(t) = ({\cal E}_G/t_G)\,[1+(t_{G}-t)/t_{0}]^{-2}/\int^{t_G}dt' [1+(t_{G}-t')/t_{0}]^{-2}$.  For $t_0\sim3 \times 10^4$~yr, the present spin-down power, $\sim 10^{34.5}$~erg~s$^{-1}$, corresponds to an upper limit on the total $e^\pm$ output of $\sim 5 \times 10^{48}$~erg (larger for smaller $t_0$~\cite{Atoyan:1995}).  Geminga's transverse velocity is $\sim200$~km~s$^{-1}$~\cite{Faherty}.  A similar radial velocity would result in a $\sim100$~pc displacement in $t_G$.

\begin{figure}[t!]
\includegraphics[width=\columnwidth,clip=true]{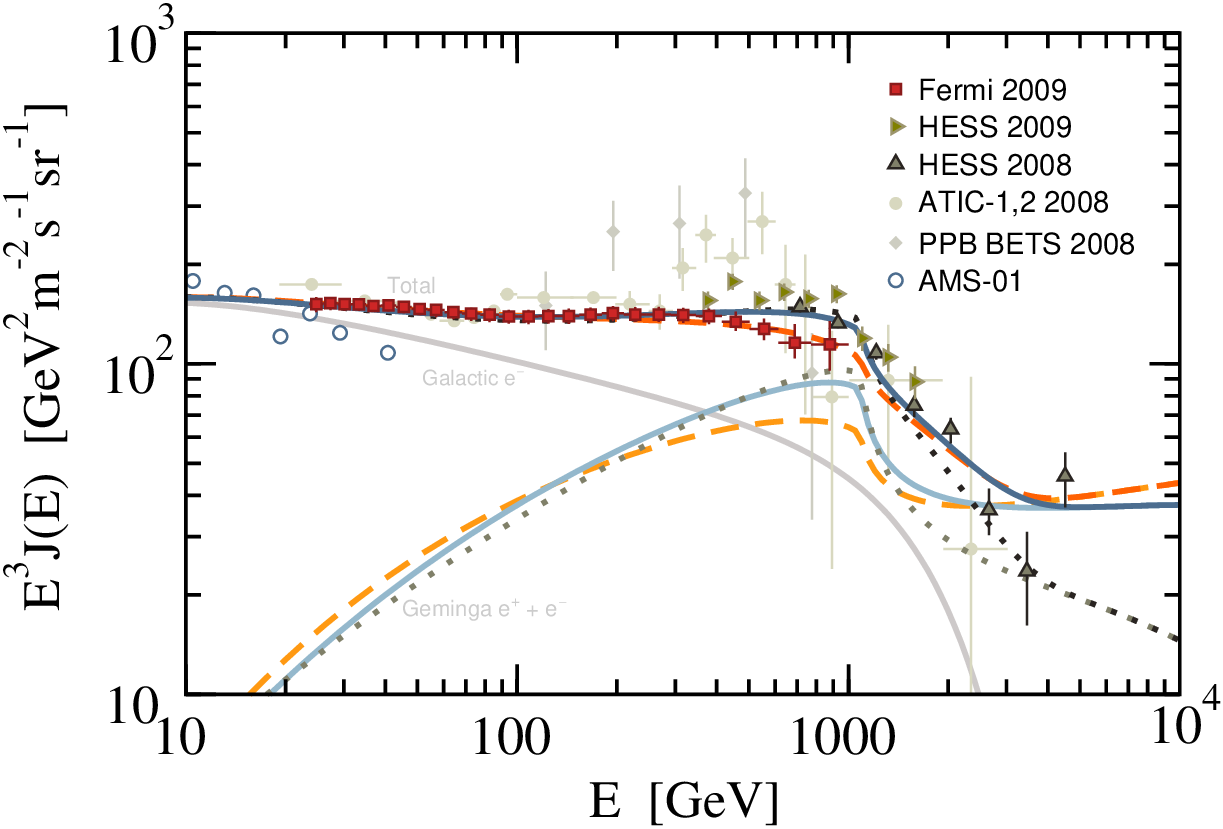}
\caption{Cosmic-ray $e^- + e^+$ data from Fermi~\cite{Fermi:2009zk}, HESS~\cite{Aharonian:2008aaa,Aharonian:2009ah}, and Refs.~\cite{Torii:2008xu,Chang:2008zzr} and AMS $e^-$-only data~\cite{Aguilar:2007yf}; with Galactic $e^-$ model modified from Ref.~\cite{Moskalenko:1997gh} (solid), contributions from Geminga (lower lines), and total (upper lines).
\label{leptons}}
\end{figure}

In Fig.~\ref{leptons}, we display the local flux of $e^- + e^+$, $J_\odot=(c/4\pi) \, n_\odot$, from our benchmark model of $\alpha=2$, within a reasonable range of parameters.  These have distances varying (from birth $\rightarrow$ present) as $r_G=150\rightarrow 250~{\rm pc},\, 220~{\rm pc},\, 250\rightarrow 200~{\rm pc}$; $e^\pm$ energy budgets of ${\cal E}_G = 1\,, 2\,, 3 \times 10^{48}$~erg; and $\delta = 0.4\,, 0.5\,, 0.6$, respectively (lower dotted, solid, dashed lines).  The energy in $e^\pm$ estimated for several younger TeV PWN are at least as large as these (e.g., \cite{deJager2007,Aharonian:2006zb,Aharonian:2006xx}).  Since the bulk of the energy is released in this early spin-down phase, the initial location is the most important.  Adding to these the primary $e^-$ spectrum of Moskalenko and Strong~\cite{Moskalenko:1997gh}, with the normalization decreased by 35\% and an added exponential cutoff at 2~TeV (in order to not exceed HESS data), yields the total $e^- + e^+$ flux (upper lines).

The spectral feature at $\sim 1$~TeV naturally results from a combination of energy losses and pulsar age and distance (see Fig.~4 of Ref.~\cite{Aharonian(1995)} for comparison).  The multi-TeV extension (beyond the last HESS point) is due to the continuous injection of particles, as evidenced by the Milagro observations today.  Combining these with the expectations for the secondary $e^\pm$ fluxes~\cite{Moskalenko:1997gh} (see also~\cite{Delahaye:2008ua}), we compare our positron fraction to measurements in Fig.~\ref{excess} (note that solar modulation may account for disagreements between data below $\sim\,$10~GeV~\cite{Adriani:2008zr,clempriv:2008}).

It is thus plausible that Geminga is the long-sought~\cite{Shen} local source of electrons and positrons, influencing the spectra measured by Fermi~\cite{Fermi:2009zk} (down to tens of GeV) and HESS~\cite{Aharonian:2008aaa,Aharonian:2009ah} in the TeV, although we emphasize that certain parameters and the underlying Galactic primary spectrum remain uncertain.  The PAMELA~\cite{Casolino:2007wr} and AMS~\cite{Aguilar:2007yf} experiments can measure the $e^-$ and $e^+$ spectra separately to isolate this component (since the $e^-$ spectrum from Geminga should be identical to the $e^+$).

\textit{Conclusions.}--- 
The discovery of high-energy gamma rays from an extended region around Geminga by Milagro reveals the presence of $\gtrsim 100$~TeV $e^\pm$, as observed indirectly within the x-ray PWN~\cite{Caraveo,Pavlov}.  A considerable amount of data should become available as new experiments examine the surrounding area.  This will help in developing more detailed models that account for both time and spatial evolution in the $e^\pm$ spectra, directly coupled to cosmic-ray propagation~\cite{Aharonian:2004yt}.  One need is a better-determined distance, the most recent quoted being $r_{G}\sim 250^{+120}_{-62}\,$pc~{\cite{Faherty}.  We briefly discuss implications for several categories of experiments.

\textit{Fermi:}
While the observed features of Geminga will depend upon details such as whether the source is roughly spherical or preferentially oriented, we would generally expect the source to become ``larger'' with decreasing energy, reflecting the decrease in IC cooling time with energy.  Our inspection of the point-source subtracted sky map from EGRET~\cite{EGRET} indicates emission in the GeV range of a size comparable to the Milagro source.  Fermi~\cite{Gehrels:1999ri} should be able to more effectively separate the bright pulsed signal to study diffuse emission.

\textit{TeV gamma rays:}
Obtaining a detailed spectrum and morphology of the source in the TeV regime will be vital for further interpretation of the nature of the particles present.  Already, VERITAS~\cite{Weekes:2001pd} has placed rather-tight upper limits on a point source at the location of Geminga~\cite{Maier:2008vw}.  Further study of the expected extended source is needed to better estimate the total energetics.  In HESS J1825--137, the surface brightness was seen to drop off as $\sim 1/\theta$, inconsistent with pure diffusion and suggestive of convection, and the gamma-ray spectrum was measured to soften with increasing distance from its pulsar~\cite{Aharonian:2006zb}.  We expect similar behavior from Geminga if the same mechanisms are at work, the latter of which would be a distinct signature of $e^\pm$ cooling~\cite{Aharonian:2006zb}.  Also, studying the extended TeV emission from an old, radio-quiet neutron star should have implications for some heretofore unidentified TeV sources.

\textit{Electrons and Positrons:}
Due to the spin down of the pulsar, it is possible that the Geminga source was much brighter in the past and dominated the TeV sky.  It is from this time that multi-GeV $e^\pm$ may still be reaching us today.  If Geminga does account for a substantial fraction of the total $e^- + e^+$ spectrum, a mild anisotropy may be present~\cite{Buesching:2008hr,Hooper:2008kg}.  Since the distance to Geminga does not greatly exceed the scale for field fluctuations of $\sim 100$~pc~\cite{Strong:2007nh}, with detailed multi-wavelength studies, local diffusion parameters might be determined (which may differ from global values estimated across the Galaxy).  Additionally, as Geminga remains a source of $\sim 100$~TeV $e^\pm$, it may result in a $> 10$~TeV lepton flux at Earth.

\textit{Neutrinos:}
The gamma rays might be produced via the decay of neutral pions produced in hadronic scattering~\cite{Gaisser:1990vg} if a nucleonic wind carries away the spin-down energy (as proposed~\cite{Horns2006} for Vela~X; but see Ref.~\cite{deJager2007}).  However, this requires $\sim 10^{47-48}$~erg of protons to be present and confined for $\gtrsim 10^{5}$~yr (likely confining $e^\pm$ as well), disfavoring this scenario.  There would also be a neutrino flux from charged pions.  This would be similar to MGRO J2019+37~\cite{Abdo:2006fq}, so that IceCube~\cite{Ahrens:2002dv} would expect $\gtrsim\,$1 (0.2) neutrino-induced muon per year with energy $>1$ (10)~TeV (see \cite{Beacom:2007yu}).  An improved measurement of the source extent is needed to estimate the atmospheric background, which can exceed this rate for a radius much larger than a degree~\cite{Ahrens:2002dv}.  Even if no neutrinos are found, IceCube will be able to provide valuable constraints~\cite{Abbasi:2008ih}.

\vspace*{0.7cm}
%
We thank John Beacom, Jim Beatty, John Clem, Tom Gaisser, Francis Halzen, Jamie Holder, Carsten Rott, and Todd Thompson for discussions; Felix Aharonian, Andrew Strong, and our Referees for comments.
HY and TS are supported by DOE Grant DE-FG02-91ER40626; MDK by DOE Grant DE-FG02-91ER40690.



\begin{thebibliography}{99}

\bibitem{Bignami:1996tp}
  G.~F.~Bignami and P.~A.~Caraveo,
  Ann.\ Rev.\ Astron.\ Astrophys.\  {\bf 34}, 331 (1996).
  
\bibitem{Abdo:2007ad}
  A.~A.~Abdo {\it et al.},
  Astrophys.\ J.\  {\bf 664}, L91 (2007).

\bibitem{Abdo:2009ku}
  A.~A.~Abdo {\it et al.},
  arXiv:0904.1018.

\bibitem{Aharonian:2006zb}
  F.~Aharonian {\it et al.},
  Astron.\ Astrophys.\  {\bf 460}, 365 (2006).

\bibitem{Aharonian:2006xx}
  F.~Aharonian {\it et al.},
  Astron.\ Astrophys.\  {\bf 448}, L43 (2006).
  
\bibitem{Barwick:1997ig}
  S.~W.~Barwick {\it et al.},
  Astrophys.\ J.\  {\bf 482}, L191 (1997).
  
\bibitem{Beatty:2004cy}
  J.~J.~Beatty {\it et al.},
  Phys.\ Rev.\ Lett.\  {\bf 93}, 241102 (2004).
  
\bibitem{Adriani:2008zr}
  O.~Adriani {\it et al.},
  Nature, {\bf 458}, 607 (2009).

\bibitem{Moskalenko:1997gh}
  I.~V.~Moskalenko and A.~W.~Strong,
  Astrophys.\ J.\  {\bf 493}, 694 (1998).

\bibitem{Coutu:1999ws}
  S.~Coutu {\it et al.},
  Astropart.\ Phys.\  {\bf 11}, 429 (1999).

\bibitem{Aharonian(1995)}
  F.~Aharonian {\it et al.},
  Astron.\ Astrophys.\  {\bf 294}, L41 (1995).






\bibitem{DM}
%
  J. Silk and M. Srednicki,
  Phys.\ Rev.\ Lett.\  {\bf 53}, 624 (1984);
%
  A. Tylka,
  Phys.\ Rev.\ Lett.\  {\bf 63}, 840 (1989);
%
  S. Rudaz and F. W. Stecker,
  Astrophys.\ J.\  {\bf 325}, 16 (1988);
%
  M. Kamionkowski and M. S. Turner,
  Phys.\ Rev.\  D {\bf 43}, 1774 (1991);
%
  E. A. Baltz and J. Edsjo,
  Phys.\ Rev.\  D {\bf 59}, 023511 (1998);
%
  J. L. Feng {\it et al.},
  Phys.\ Rev.\  D {\bf 63}, 045024 (2001);
%
  D. Hooper and J. Silk,
  Phys.\ Rev.\  D {\bf 71}, 083503 (2005).

\bibitem{Goldreich:1969sb}
  P.~Goldreich and W.~H.~Julian,
  Astrophys.\ J.\  {\bf 157}, 869 (1969).

\bibitem{Rees:1974nr}
  M.~J.~Rees and J.~E.~Gunn,
  Mon.\ Not.\ Roy.\ Astron.\ Soc.\  {\bf 167}, 1 (1974);
  C.~F.~Kennel and F.~V.~Coroniti,
  Astrophys.\ J.\  {\bf 283}, 694 (1984).

\bibitem{Gaensler:2006ua}
  B.~M.~Gaensler and P.~O.~Slane,
  Ann.\ Rev.\ Astron.\ Astrophys.\  {\bf 44}, 17 (2006);
  J.~Arons,
  arXiv:0708.1050.

\bibitem{Fermi:2009zk}
  A. A. Abdo {\it et al.}
  Phys.\ Rev.\ Lett.\  {\bf 102}, 181101 (2009).

\bibitem{Aharonian:2008aaa}
  F.~Aharonian {\it et al.},
  Phys.\ Rev.\ Lett.\  {\bf 101}, 261104 (2008).

\bibitem{Aharonian:2009ah}
  F.~Aharonian {\it et al.},
  arXiv:0905.0105.

\bibitem{Taylor:1993ba}
  J.~H.~Taylor {\it et al.},
  Astrophys.\ J.\ Suppl.\  {\bf 88}, 529 (1993).

\bibitem{Caraveo}
  P.~A.~Caraveo {\it et al.},
  Science {\bf 301}, 1345 (2003).

\bibitem{Pavlov}
  G.~G.~Pavlov {\it et al.},
  Astrophys.\ J.\  {\bf 643}, 1146 (2006).
  
\bibitem{Blumenthal:1970gc}
  G.~R.~Blumenthal and R.~J.~Gould,
  Rev.\ Mod.\ Phys.\  {\bf 42}, 237 (1970).

\bibitem{deJager2007}
  O.~C.~de~Jager,
  Astrophys.\ J.\  {\bf 658}, 1177 (2007).

\bibitem{Zhang:1999ua}
  B.~Zhang and A.~K.~Harding,
  Astrophys.\ J.\  {\bf 532}, 1150 (2000);
  J.~A.~Hibschman and J.~Arons,
  Astrophys.\ J.\  {\bf 560}, 871 (2001).

\bibitem{Atoyan:1995}
  A.~M.~Atoyan {\it et al.},
  Phys.\ Rev.\  D {\bf 52}, 3265, (1995).

\bibitem{Strong:2007nh}
  A. W. Strong {\it et al.},
  Ann.\ Rev.\ Nucl.\ Part.\ Sci.\  {\bf 57}, 285 (2007).

\bibitem{Aloisio:2008tx}
  R.~Aloisio {\it et al.},
  Astrophys.\ J.\  {\bf 693}, 1275 (2009).

\bibitem{Gunn:1969ej}
  J.~E.~Gunn and J.~P.~Ostriker,
  Nature  {\bf 221}, 454 (1969).


\bibitem{Faherty}
  J.~Faherty {\it et al.},
  Astrophys.\ Space Sci.\ {\bf 308}, 225 (2007).
  
\bibitem{Delahaye:2008ua}
  T.~Delahaye {\it et al.},
  arXiv:0809.5268.

\bibitem{clempriv:2008}
  J.~Clem and P.~Evenson,
  Proc.\ 30th Intl.\ Cosmic Ray Conf.\  {\bf 1}, 477 (2008).






\bibitem{Shen}
  C.~S.~Shen,
  Astrophys.\ J.\  {\bf 162}, L181 (1970).

\bibitem{Casolino:2007wr}
  M.~Casolino {\it et al.},
  Adv.\ Space Res.\  {\bf 42}, 455 (2008).

  
  
\bibitem{Aguilar:2007yf}
  M.~Aguilar {\it et al.},
  Phys.\ Lett.\  B {\bf 646}, 145 (2007).


\bibitem{Aharonian:2004yt}
  F.~A.~Aharonian, \textit{Very High Energy Cosmic Gamma Radiation},
  (World Scientific, 2004).

\bibitem{EGRET}
  A.~Cillis and R.~Hartman,
  Astrophys.\ J.\  {\bf 621}, 291 (2005).
  %

\bibitem{Gehrels:1999ri}
  N.~Gehrels and P.~Michelson,
  Astropart.\ Phys.\  {\bf 11}, 277 (1999).
  
\bibitem{Weekes:2001pd}
  T.~C.~Weekes {\it et al.},
  Astropart.\ Phys.\  {\bf 17}, 221 (2002).
  
\bibitem{Maier:2008vw}
  G.~Maier [VERITAS collaboration],
  arXiv:0810.0515.

\bibitem{Buesching:2008hr}
  I.~Buesching {\it et al.},
  Astrophys.\ J.\  {\bf 678}, L39 (2008).

\bibitem{Hooper:2008kg}
  D.~Hooper {\it et al.},
  JCAP {\bf 0901}, 025 (2009).

\bibitem{Gaisser:1990vg}
  T.~K. Gaisser, \textit{Cosmic Rays and Particle Physics},
  (Cambridge Univ. Press, Cambridge, 1990).


\bibitem{Horns2006}
  D.~Horns {\it et al.},
  Astron.\ Astrophys.\  {\bf 451}, L51 (2006).

\bibitem{Abdo:2006fq}
  A.~A.~Abdo {\it et al.},
  Astrophys.\ J.\  {\bf 658}, L33 (2007).

\bibitem{Ahrens:2002dv}
  J.~Ahrens {\it et al.},
  Nucl.\ Phys.\ Proc.\ Suppl.\  {\bf 118}, 388 (2003);
  J.~Ahrens {\it et al.},
  Astropart.\ Phys.\  {\bf 20}, 507 (2004).

\bibitem{Beacom:2007yu}
  J.~F.~Beacom and M.~D.~Kistler,
  Phys.\ Rev.\  D {\bf 75}, 083001 (2007);
  L.~Anchordoqui {\it et al.},
  Phys.\ Rev.\  D {\bf 76}, 067301 (2007);
  M.~D.~Kistler and J.~F.~Beacom,
  Phys.\ Rev.\ D {\bf 74}, 063007 (2006).



\bibitem{Abbasi:2008ih}
  R.~Abbasi {\it et al.},
  Phys.\ Rev.\  D {\bf 79}, 062001 (2009).


\bibitem{Torii:2008xu}
  S.~Torii {\it et al.},
  arXiv:0809.0760.

\bibitem{Chang:2008zzr}
  J.~Chang {\it et al.},
  Nature {\bf 456}, 362 (2008).




\end{thebibliography}
\end{document}